\documentclass[graybox,natbib,nosecnum]{svmult}
\bibpunct{(}{)}{;}{a}{}{,} 

\pdfoutput=1 

\usepackage{comment}
\usepackage{amsmath}
\usepackage{mathptmx}       
\usepackage{helvet}         
\usepackage{courier}        
\usepackage{type1cm}        

\usepackage{makeidx}         
\usepackage{graphicx}        
\usepackage{multicol}        
\usepackage[bottom]{footmisc}
\usepackage[normalem]{ulem}	
\usepackage{hyperref}  
\usepackage{url}  

\usepackage{soul}

\makeindex  
\begin{document}

\def\ltsima{$\; \buildrel < \over \sim \;$}
\def\lsim{\lower.5ex\hbox{\ltsima}}
\def\gtsima{$\; \buildrel > \over \sim \;$}
\def\gsim{\lower.5ex\hbox{\gtsima}}

\title*{The Transiting Exoplanet Survey Satellite}
\titlerunning{The Transiting Exoplanet Survey Satellite}

\author{Joshua N.\ Winn}
\authorrunning{J.\ N.\ Winn}
\institute{Princeton University, Princeton, NJ, USA \email{jnwinn@princeton.edu}}

\maketitle

\vskip -1.5in
\abstract{A transiting planet invites us to measure its
size, mass, orbital parameters, atmospheric
composition, and other characteristics.
But the invitation can only be accepted if the host star is bright
enough for precise measurements of its flux and spectrum.
NASA's Transiting Exoplanet Survey Satellite (TESS) is dedicated to finding such favorable systems. Operating from a 13.7-day elliptical orbit around the Earth, TESS uses four 10.5\,cm telescopes to capture optical images of a $24^\circ$ by $96^\circ$ field of view. By shifting the field of view every 27 days, TESS can survey most of the sky every few years.
In its first six years, TESS has identified approximately 7{,}000 planet candidates, with several hundred confirmed as planets. Mass measurements of these planets allow astronomers to differentiate between rocky ``super-Earths'' and gas-rich or volatile-rich ``mini-Neptunes,'' while observations with the James Webb Space Telescope are revealing the secrets of their atmospheres.
TESS has discovered planets orbiting a wide range of stars, including young stars, low-mass stars, binary stars, and even a white dwarf star. Beyond planet detection, TESS probes the optical variability of stars and a diverse array of other astronomical objects, including asteroids, comets, supernovae, and active galactic nuclei.}

\section{A brief history of transit surveys}
\label{sec:history_of_transit_surveys}

Among the various methods for detecting exoplanets, the transit method is the simplest to explain.
Even young children can grasp the concept:
when a planet passes directly in front of
a star, the star's light dims slightly.
However, it is not child's play to monitor a star's brightness precisely enough to detect a planetary transit,
nor to monitor enough stars to overcome the low
likelihood that any given planet's orbit 
is closely aligned with our line of sight.
Even when hunting for a ``hot Jupiter'' --- the easiest kind of planet  to detect --- astronomers must monitor thousands of stars over several weeks, waiting for 
one of them to fade by just 1\% for a few hours.

\begin{figure}
\centerline{\includegraphics[scale=0.5]{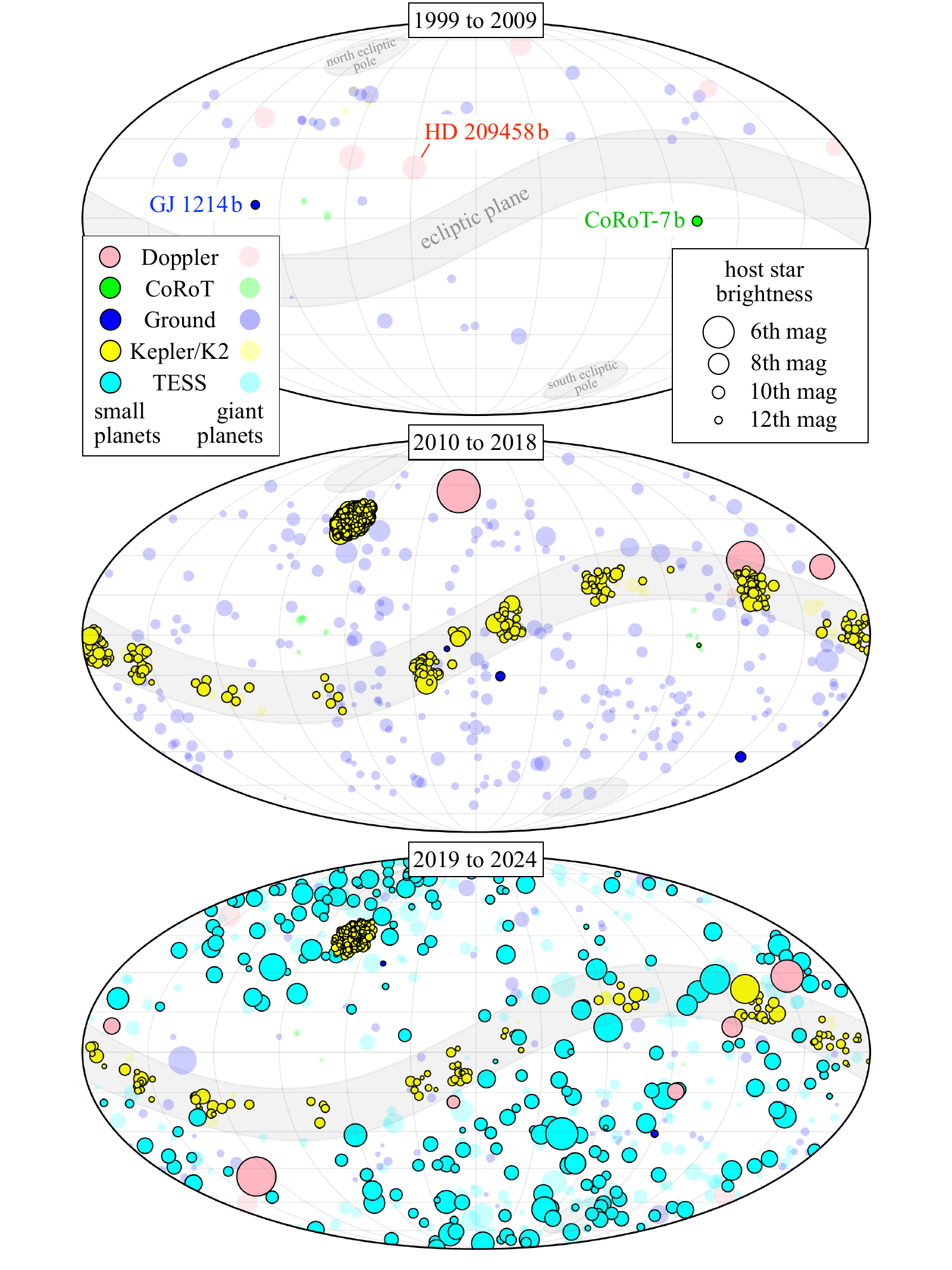}}
\caption{{\bf Three eras of transiting planets.}
Each of these three equal-area sky projections shows the equatorial coordinates of transiting planets
discovered in a particular era. The ecliptic plane and poles are
shaded gray.
The color of each point conveys the type of survey
responsible for the planet's discovery.
Translucent circles represent planets larger than Neptune (3.86~$R_\oplus$) and opaque circles represent smaller planets.
The size of each circle encodes
the star's apparent magnitude in the TESS bandpass.
\label{fig:skymaps}
}
\end{figure}

The first known transiting planet, HD\,209458\,b, was not
discovered with the transit method.
Instead, the discoverers ``cheated'' by monitoring a handful
of stars that already showed evidence of a hot
Jupiter from the Doppler method \citep{Charbonneau+2000,Henry+2000}.
The Doppler data enabled them to predict when
transits would occur if the
planet's orbit were nearly
aligned with our line of sight.
The transiting planets discovered by this
Doppler-first method are marked in red
on the sky maps in Figure~\ref{fig:skymaps}.
Other notable planets found with this method
include the transiting ``super-Neptune'' GJ\,436\,b \citep{Gillon+2007},
the ``eccentric giant'' HD\,80606\,b \citep{Moutou+2009,Fossey+2009,GarciaMelendoMcCullough2009},
and the transiting ``super-Earth'' 55\,Cnc\,e \citep{Winn+2011,Demory+2011}.

Between 2003 and 2009, dozens of transiting planets were discovered in ground-based transit surveys that monitored as many stars as possible without any prior knowledge of planets. These discoveries
are marked with dark blue points in Figure~\ref{fig:skymaps}; among the earliest were OGLE-TR-56\,b \citep{Konacki+2003} and TrES-1\,b \citep{Alonso+2004}. 
Many well-known exoplanets
bear the names of the surveys from
this era, particularly HAT \citep{Bakos2018}
and WASP \citep{Pollacco+2006}.
Nearly all the planets found in these ground-based surveys were hot Jupiters. Detecting smaller planets was more challenging due to the limitations on photometric precision imposed by Earth's fluctuating atmosphere. Detecting planets with wider orbits was difficult for two reasons: the geometric probability of transits decreases with orbital distance, and longer orbital periods mean fewer transits, making them easy to miss altogether if they occur during the daytime or spells of bad weather.

The pioneering transit surveyors also learned
that a sequence of brightness dips in a star's light
curve is not necessarily due to a transiting planet.
Often, the ``star''
turned out to be an unresolved combination of a bright
star and a faint eclipsing binary.
Distinguishing those cases from
genuine planetary transits 
proved challenging.
To confirm a transit candidate as a true planet, the astronomical community expected a measurement of
the planet's mass using the Doppler method,
which required a high-resolution spectrograph and a much larger
telescope than was needed to detect the photometric signal.
Surveys that had monitored bright stars across wide
fields using small-aperture telescopes
had an easier time confirming planets
compared to those that focused on
fainter stars within the narrower fields of view of large-aperture telescopes.

The key takeaway from these early surveys --- {\it brighter is better} ---
was the driving force behind the creation of 
the Transiting Exoplanet Survey Satellite (TESS).
The most valuable transiting
planets are those orbiting the brightest
stars. Bright stars not only ease the process
of confirming planet candidates,
but also enable a wide range of
follow-up observations to measure the planets' masses, orbital parameters,
and atmospheric spectra. In many ways, TESS
is the space-based sequel to the
small-aperture, wide-field transit surveys of this era.

Space telescopes provide significant advantages for transit surveys. In space, brightness measurements are far more precise and can be made almost continuously, regardless of the time of day or weather back on Earth's surface.  The main drawbacks of space telescopes are their extravagant costs and lengthy germination times. In fact, long before the lesson {\it brighter is better} had sunk in, two other space-based transit surveys
had already been planned: CoRoT and Kepler.

The CoRoT mission was first proposed to the French national space agency in 1993, and launched in 2006. Conceived as a stellar astrophysics investigation, its original name was CoRot, a blend of Convection and Rotation. After exoplanets became a hot topic, the 
mission's goals expanded to include them
and the final `t' was elevated to `T' for Transits.
From its low-Earth orbit, CoRoT used a 30-cm optical telescope to conduct precise photometry for extended periods, often months at a time. Due to the survey camera's small field of view (a few degrees across), CoRoT had to monitor relatively faint stars to assemble a large enough sample for a meaningful transit survey. As a result, the mission faced significant challenges in detecting and confirming planets. Nevertheless, CoRoT successfully identified several dozen hot Jupiters and discovered CoRoT-7\,b, the first known transiting super-Earth orbiting a Sun-like star (see the green points in Figure 1; \citealt{Deleuil+2018}).

The Kepler mission was first proposed to NASA in 1992 and launched in 2009, igniting a revolution in exoplanetary science. Kepler discovered several thousand transiting planets (the yellow points in Figure~\ref{fig:skymaps}), providing an invaluable dataset for statistical studies and uncovering entirely new categories of exoplanets. The mission's accomplishments, too numerous to detail here, were summarized by \cite{Borucki2016}. Importantly,
Kepler was not an outgrowth of the earlier transit survey efforts. Conceived before the discovery of any exoplanets, Kepler was designed to find Earth-sized planets within the habitable zones of Sun-like stars. Between 2009 and 2013, Kepler used a meter-class telescope to monitor several hundred thousand stars within a
$10^\circ$ by $10^\circ$ field of view. However, most of these stars were too faint for Doppler spectroscopy to be practical, making it impossible to confirm most of the planets. Instead, the scientific community came to accept transiting
planets as ``validated'' if the data passed a series of statistical tests to guard against false positives \citep{Batalha+2010}.

During observations, the Kepler telescope maintained a fixed pointing direction using four ``reaction wheels'' spinning at thousands of revolutions per minute. Adjusting their spin rates caused the spacecraft to rotate, via the conservation of angular momentum.
However, after two reaction wheels failed, Kepler could no longer maintain stable pointing at the same star field it had been monitoring for four years. The only part of the sky
where stable pointing could be achieved, using the two remaining reaction wheels and onboard thrusters, was the ecliptic plane.
Thus, in 2014, the mission was rebranded ``K2'' and the telescope spent the next four years examining
a series of star fields along the ecliptic plane \citep{Howell+2014}.
Together, the Kepler and K2 surveys covered approximately 6\% of the celestial sphere before the spacecraft ran out of fuel and ceased operations in 2018.

The TESS mission was launched that same year.
The timing makes it seem as though TESS was the designated
successor of the Kepler mission;
however, while chronologically accurate, this portrayal obscures the differences between the two missions.
Kepler's primary goal was to conduct a planetary census, tallying Earth-like planets, rather than facilitating follow-up observations of individual planets. In contrast, TESS was designed with follow-up observations as the primary consideration.
TESS plucks the lowest-hanging
fruit from the exoplanet orchard: short-period planets, for which transits are geometrically
more likely and occur more frequently in time.
Compared to Kepler, TESS monitors brighter stars over wider fields of view (see the cyan points in the lowest panel of Figure~\ref{fig:skymaps}). TESS also surveys a much larger number of stars, over broader ranges of mass and age.
However, TESS typically observes any given star for only one or two months at a time, much shorter than Kepler's original
four-year campaign. The rest of this article describes why TESS was conceived, how it operates, and what it
has achieved so far.

\section{A brief history of TESS}
\label{sec:history_of_tess}

\begin{figure}[t!]
\centerline{\includegraphics[width=11.75cm]{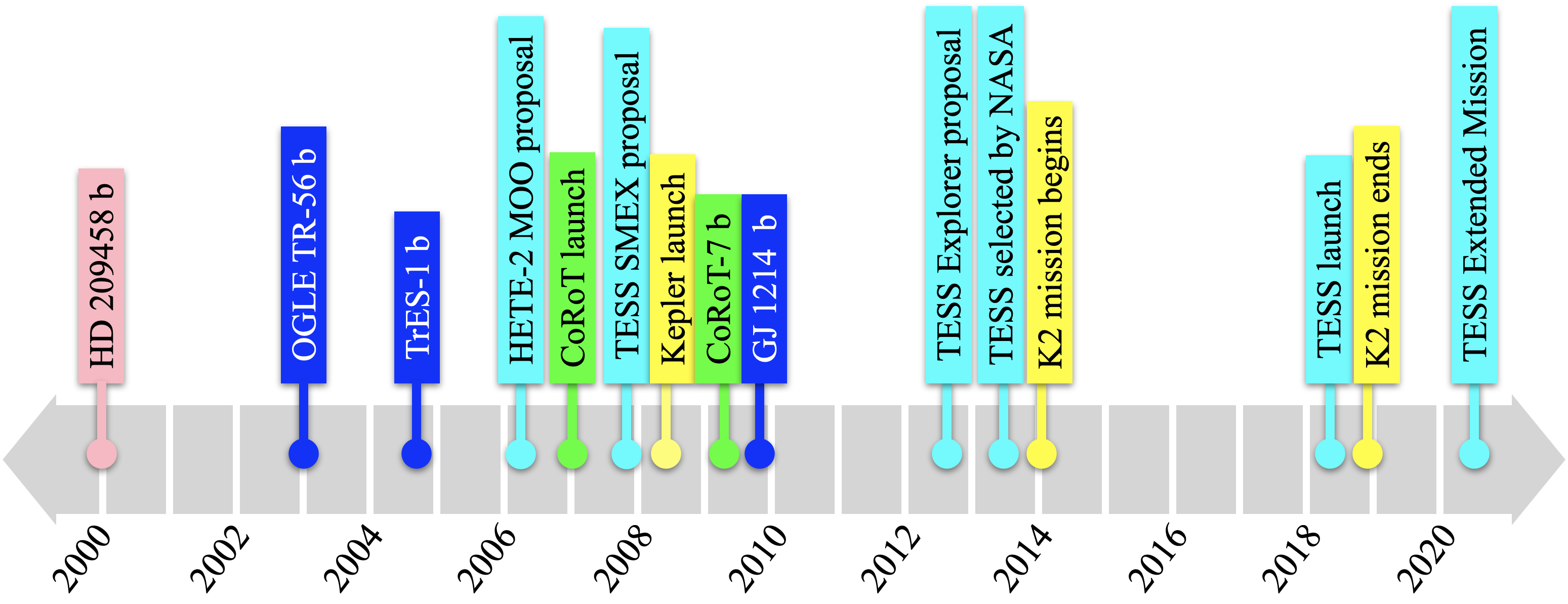}}
\caption{{\bf Timeline of transit surveys.}
Some key planet discoveries are marked, along with milestones
for the CoRoT, Kepler, and TESS space missions.
There are also plans for two future space missions to
detect transiting planets:
PLATO \citep{Rauer+2024}, a European mission scheduled for launch in
2026; and Earth Two \citep{Ge+2022}, a Chinese mission aiming for launch in 2028.
\label{fig:timeline}
}
\end{figure}

In 2006, George Ricker from the Massachusetts Institute of Technology (MIT) led a
NASA proposal to
conduct a space-based survey for transiting hot Jupiters
(see the timeline in Figure~\ref{fig:timeline}).
This ``mission of opportunity'' had a relatively low cost
because instead of building and launching
a new space telescope, he and his team would repurpose two
optical star trackers already in orbit
aboard the High-Energy Transient
Explorer 2, a spacecraft designed to study gamma-ray bursts.
Although this proposal was not selected for funding, it nucleated the group that would eventually become the TESS team, including David Latham and others from the Harvard-Smithsonian Center for Astrophysics. Latham, a co-investigator in the Kepler mission and a spectroscopist familiar with the challenges of confirming planet candidates from ground-based surveys, 
Latham was well placed to appreciate the importance of a
standalone space mission that would focus on bright stars.

Harvard was also the headquarters for an innovative transit survey called MEarth, led by David Charbonneau. By concentrating on small, red, low-luminosity stars, the MEarth survey was more sensitive to smaller planets and habitable-zone planets than other contemporary surveys \citep{BertaThompson+2013}. (A similar strategy was later employed by \citealt{Gillon+2017} to discover the extraordinary TRAPPIST-1 system, which contains seven transiting Earth-sized planets.) The lesson that ``redder is better'' influenced the design of the TESS mission, particularly the emphasis on observing at far-red wavelengths.

Ricker and Latham initially sought private funding for the mission, but soon another opportunity arose to propose a NASA mission. They assembled a team from their institutions, along with collaborators from NASA's Ames Research Center and other organizations. The 2008 proposal for a NASA Small Explorer Mission (approximately \$100 million) featured a spacecraft equipped with six 13-cm telescopes designed to conduct a two-year, all-sky transit survey from a low Earth orbit. However, this mission concept was not selected.

The TESS team regrouped and prepared for another opportunity to propose a mission, which did not arise until 2011. This delay allowed the team to benefit from the Kepler mission's discovery of a high abundance of close-orbiting, Earth-to-Neptune-sized planets, further strengthening the motivation for TESS. 
Another motivating factor was that
the James Webb Space Telescope (JWST) had
survived an attempted cancellation and was
scheduled for launch within a decade.
One of JWST's most anticipated capabilities was performing
spectroscopy of the atmospheres of transiting planets, especially habitable-zone planets around low-mass stars.
This made it urgent to discover the most
observationally favorable systems.

The delay also prompted the team to rethink TESS and reconfigure it as an Explorer-class mission (approximately \$200 million). One significant change was an upgrade
from a low-Earth orbit to a much higher-altitude orbit. This higher orbit allows for nearly continuous observations and bestows other advantages that are described below. However, it also increased the mission's cost and complexity
by requiring the spacecraft
to be equipped with its own propulsion system.
To stay within budget, the number of telescopes was reduced from six to four.

\begin{figure}[b]
\centerline{\includegraphics[width=11.75cm]{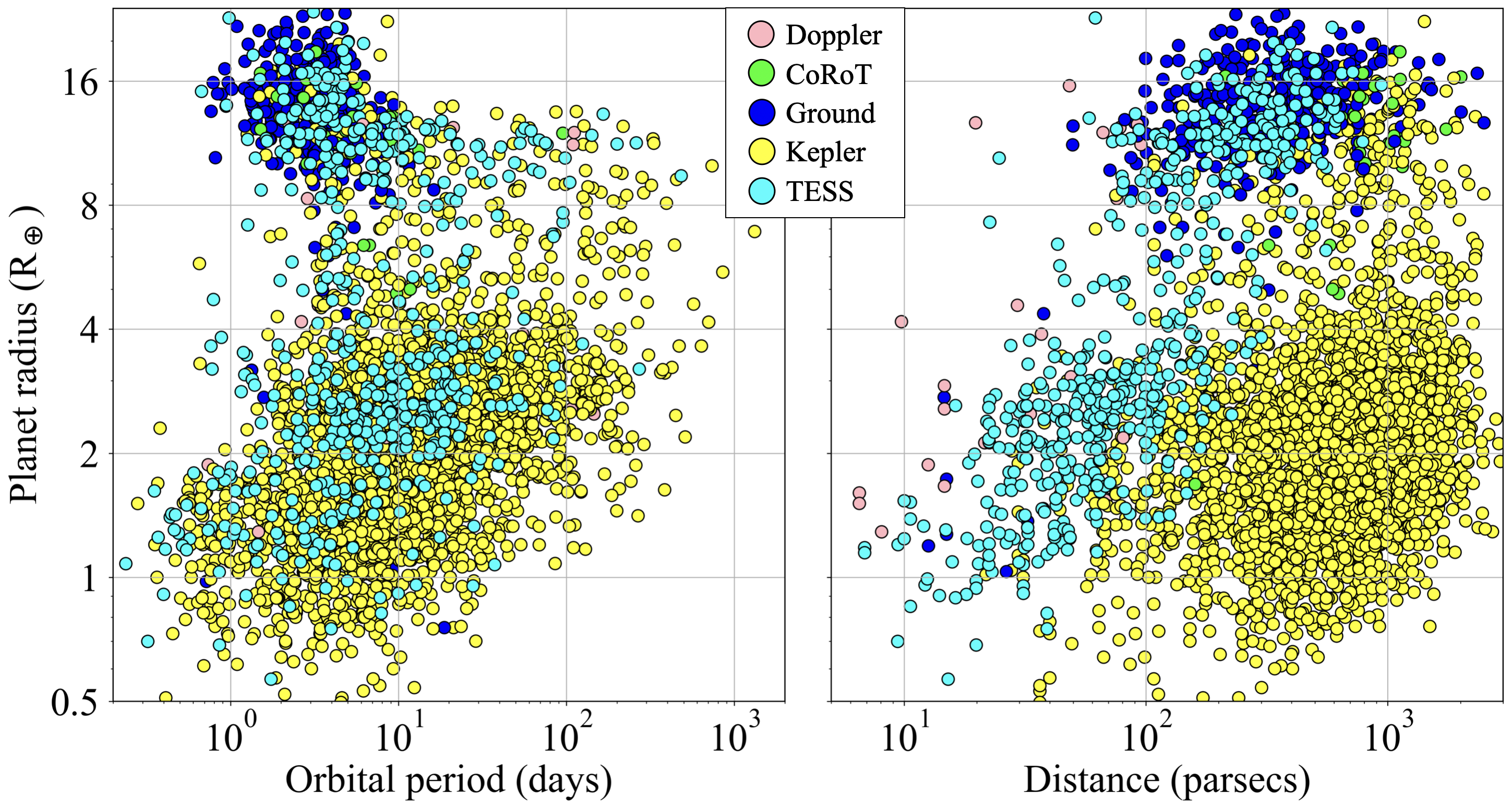}}
\caption{
{\bf TESS and other transit surveys}. {\it Left.}---Planet radius vs.\ orbital period. Compared
to ground-based surveys (dark blue points), space-based surveys
are capable of finding smaller planets. TESS planets (cyan)
are more strongly biased toward short periods than Kepler
planets (yellow). {\it Right.}---Planet radius vs.\ distance from the Solar System.
TESS is completing the census
of nearby hot Jupiters
that was begun by the
ground-based surveys
(top of the chart),
and
discovering the nearest
examples of transiting 
planets smaller than Neptune (bottom of the chart).
\label{fig:pr}
}
\end{figure}

In this new incarnation, TESS was selected by NASA in April 2013. The cameras and detectors were built by MIT's Lincoln Laboratory and Kavli Institute for Astrophysics \& Space Research, while the spacecraft was constructed by Orbital Sciences Corporation (now part of Northrop Grumman). Prior to launch, the mission was summarized by \cite{Ricker+2015}, and simulations of TESS's planet detections were presented by \cite{Sullivan+2015}. TESS was launched atop a SpaceX Falcon-9 rocket on April 18, 2018, from Cape Canaveral, Florida, and began science operations that summer.

The ``prime mission'' of TESS was to scan 85\% of the sky, detect at least 50 planets smaller than Neptune, and measure their masses using the Doppler technique. Thanks to the efforts of hundreds of astronomers involved in the TESS Follow-up Observing Program, this goal was achieved in 2021. Since then, TESS has had a series of ``extended missions'' with broader scientific objectives, expanding the sky coverage to nearly 100\% and re-observing most regions of the sky multiple times. As of this writing, approximately 7{,}000 planet candidates have been identified using TESS data. Of these, about 550 planets have been documented in the literature, several hundred have been confirmed through Doppler measurements (see Figure~\ref{fig:mr}), and about 2{,}500 scientific papers have been published based on TESS data.


\section{How TESS works}
\label{sec:spacecraft_and_orbit}

\begin{figure}[b!]
\centerline{\includegraphics[width=11.75cm]{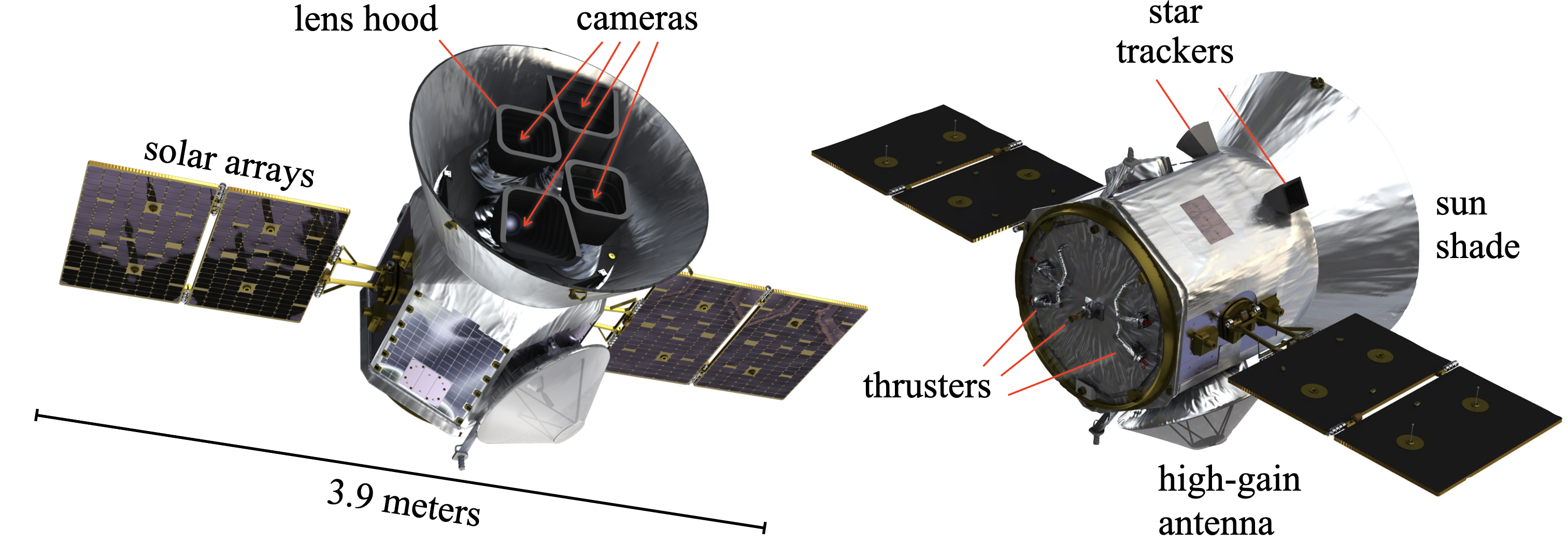}}
\caption{
{\bf Two views of TESS.} Not shown in these diagrams
are the onboard computer and the four reaction wheels, which
are inside the hexagonal body of the spacecraft.
\label{fig:spacecraft}
}
\end{figure}

\runinhead{The equipment}  The key components of the TESS spacecraft are
labeled in Figure~\ref{fig:spacecraft}.
From an astronomer's point of view, the most critical elements
are the four nearly identical wide-field optical cameras.
Each camera has a collecting aperture of diameter 10.5~cm that illuminates
a $4096^2$-pixel imaging array at the focal plane. Taking into account the optical throughput and quantum efficiency, the effective
collecting area is 63~cm$^2$ over the range
of wavelengths from 600 to 1050~nm. This far-red
bandpass 
was chosen with small, red, low-luminosity stars in mind.

The four TESS cameras cover a total field of view of $24^\circ$ by $96^\circ$, which is approximately 5.5\% of the celestial sphere. Each pixel subtends about 21 arcseconds on a side, resulting in images with an angular resolution on the order of one arcminute, comparable to human vision. Like the Kepler spacecraft, TESS employs reaction wheels for precise control of the spacecraft's orientation. Whenever maintaining a fixed pointing would require the reaction wheels to spin too rapidly --- due to the accumulation of angular momentum from solar radiation pressure --- the onboard thrusters are used to reorient the spacecraft, allowing the wheels to slow down. These ``momentum dumps'' are familiar to TESS aficionados, as they are sometimes associated with glitches in light curves.

\runinhead{The orbit} A low-Earth orbit is relatively easy to reach but poses problems for precise photometry, due to stray light from the Earth, variations in temperature and illumination from the Sun, and damage from charged particles trapped in the Earth's magnetic field. To mitigate these problems, TESS rocketed away from a low-Earth orbit and, with the help of a lunar flyby, arrived in a highly eccentric orbit with an average period of 13.7 days. This orbit provides a stable observing environment while remaining close enough to Earth to allow a high data transmission rate. A carefully designed feature of the TESS orbit is that the Moon alternately leads and lags the spacecraft by $90^\circ$ at each apogee. This 2:1 lunar resonance leads to the partial cancellation of the gravitational perturbations that would otherwise destabilize such a high-altitude orbit \citep{Gangestad+2013}.

\runinhead{The survey} 

\begin{figure}[ht!]
\includegraphics[width=11.75cm]{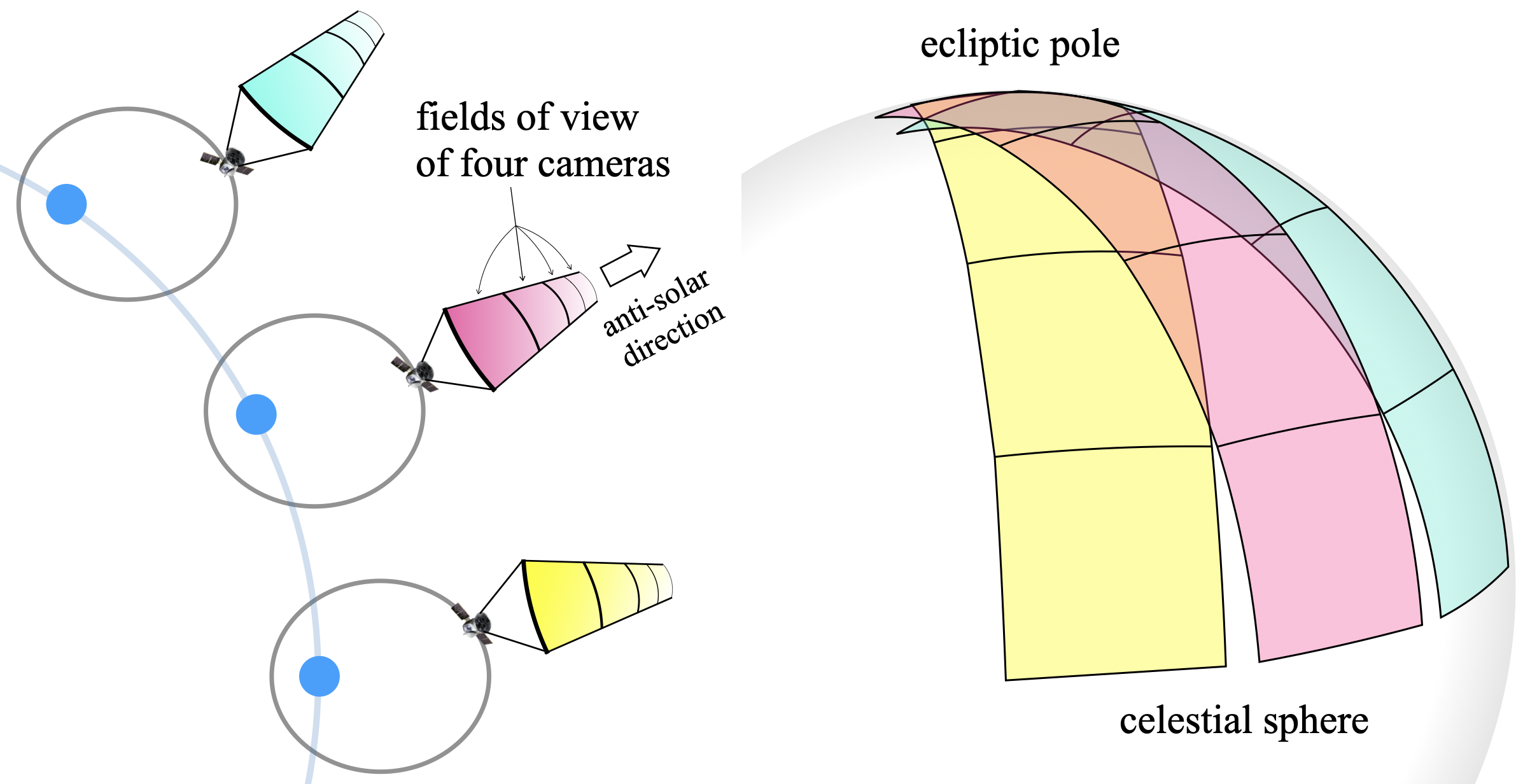}
\caption{
{\bf The TESS orbit and typical scan pattern}.
{\it Left.}---Three months in TESS's
highly elliptical orbit around the Earth.
TESS generally points away from
the Sun and observes a given field for two
spacecraft orbits (one lunar sidereal month).
The drawing is not to scale; in reality,
the perigee oscillates between about
12 and 25\,$R_\oplus$ over timescales of years.
For reference, the geosynchronous orbit is at 6.6\,$R_\oplus$
and the Moon's average orbital distance is 60\,$R_\oplus$.
The TESS orbit is inclined
by about 35$^\circ$ relative to the ecliptic plane,
reducing the frequency with which the Earth and Moon
interfere with observing and
eclipse the spacecraft's
solar panels.
{\it Right}.---A typical scan pattern, showing three rectangular
sectors that overlap at one of the ecliptic poles.
\label{fig:orbit}
}
\end{figure}

Figure~\ref{fig:orbit}
illustrates TESS's basic sky-scanning pattern.
Each ecliptic hemisphere 
is divided into 13 rectangular
sectors that overlap near the ecliptic pole.
Data are collected continuously for two consecutive spacecraft orbits, equivalent to one sidereal lunar month.
By advancing eastward to the next sector each
month, TESS stays pointed away
from the Sun. With this strategy,
a given star is observed for a duration ranging from 27 days to 351 days, depending chiefly
on the star's ecliptic latitude.

TESS has occasionally deviated from this basic plan by shifting some observing fields away from the ecliptic during periods when light from the Earth or Moon would have spoiled large portions of the data. Additionally, TESS has spent several months observing low ecliptic latitudes, aligning the long axis of its field of view with the ecliptic plane. By the end of 2024, TESS will have surveyed each hemisphere three times and covered most of the ecliptic zone at least once. For future observations, new sky-scanning patterns are being considered to strike different compromises between sky coverage and dwell time.


\section{Using TESS data}
\label{sec:data}

\runinhead{Data products}  Each TESS camera
acquires a new image every two seconds.
Since this generates more data
than can be sustainably stored and transmitted to the ground,
the onboard computer processes the
images to reduce the data volume and create two types of
data products:
\begin{enumerate}

\item {\bf Full-Frame Images} (FFIs).
Groups of consecutive images are
summed to create an image with a longer effective
exposure time.
Initially, the time between FFIs was
1800 seconds, but as the mission progressed, this cadence was shortened to 600 seconds and eventually to 200 seconds.

\item {\bf Target Pixel Files} (TPFs). For pre-selected
stars of interest, data with faster time sampling are retained. Miniature images surrounding each target star (typically $11\times 11$ pixels) are summed
over time intervals of 120 or 20 seconds.
Several thousand stars in each sector can be selected for this special treatment.
The selection was initially the responsibility of the TESS team and is now carried out by the astronomical community through the Guest Investigator program and Director's Discretionary Time.
 
\end{enumerate}

\noindent A light curve is a time series of flux measurements obtained by summing the counts in the relevant pixels of each image and subtracting the estimated contributions from the sky background and detector effects. In practice, creating light curves requires many choices to be made regarding pixel weighting schemes, background subtraction methods, and ``detrending'' --- the removal of instrumental
artifacts and
intrinsic stellar variability.
Another consideration is how to correct for ``blending'' --- 
the fact that the flux recorded by TESS is often from unresolved
combinations of multiple sources.
Consequently, TESS light curves are available in several versions, including those produced by TESS's Science Processing Operations Center (SPOC), the MIT Quick Look Pipeline (QLP), and light curves generated by external groups.

\begin{figure}[ht!]
\centerline{\includegraphics[width=12.5cm]{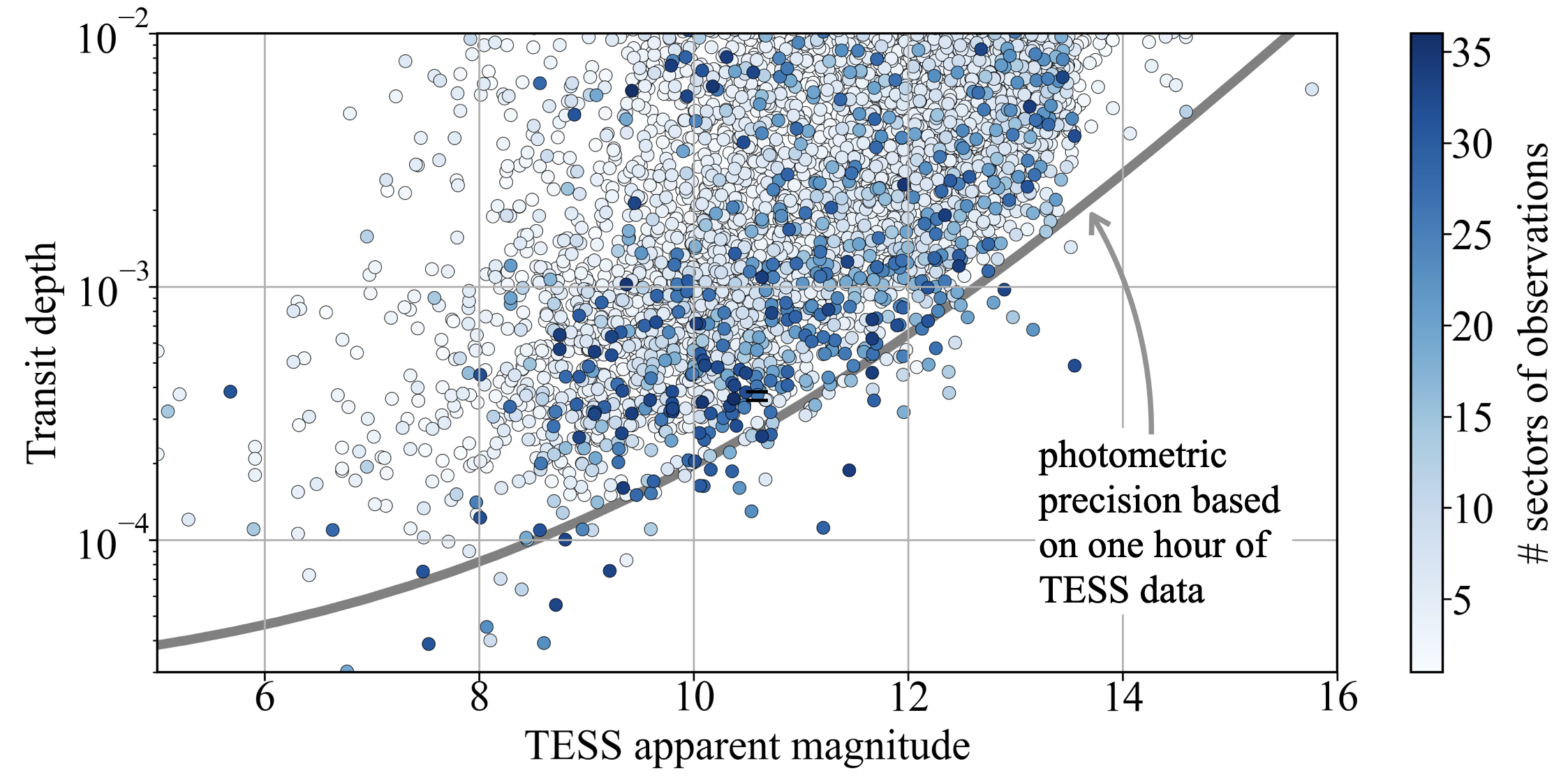}}
\caption{
{\bf Photometric performance.} 
Plotted are the transit depths
of all TESS Objects of Interest (excluding known
false positives), as a function of the apparent magnitude of the host star. The transit depth is roughly
$(r/R)^2$, where $r$ and $R$ are the radii of the planet
and star.
The gray curve is the approximate
limiting precision with which the brightness of a star can be
measured by TESS based on one hour of data \citep[see Table 3 of][]{Kunimoto+2022}.
The color of each data point conveys the number of sectors
in which the star was observed, i.e., the quantity of available TESS
data. Dark colors generally correspond to stars near the ecliptic poles,
which are included in many sectors, enabling more sensitive
searches for small planets.
\label{fig:tois}
}
\end{figure}

\begin{figure}[ht!]
\centerline{\includegraphics[width=11.75cm]{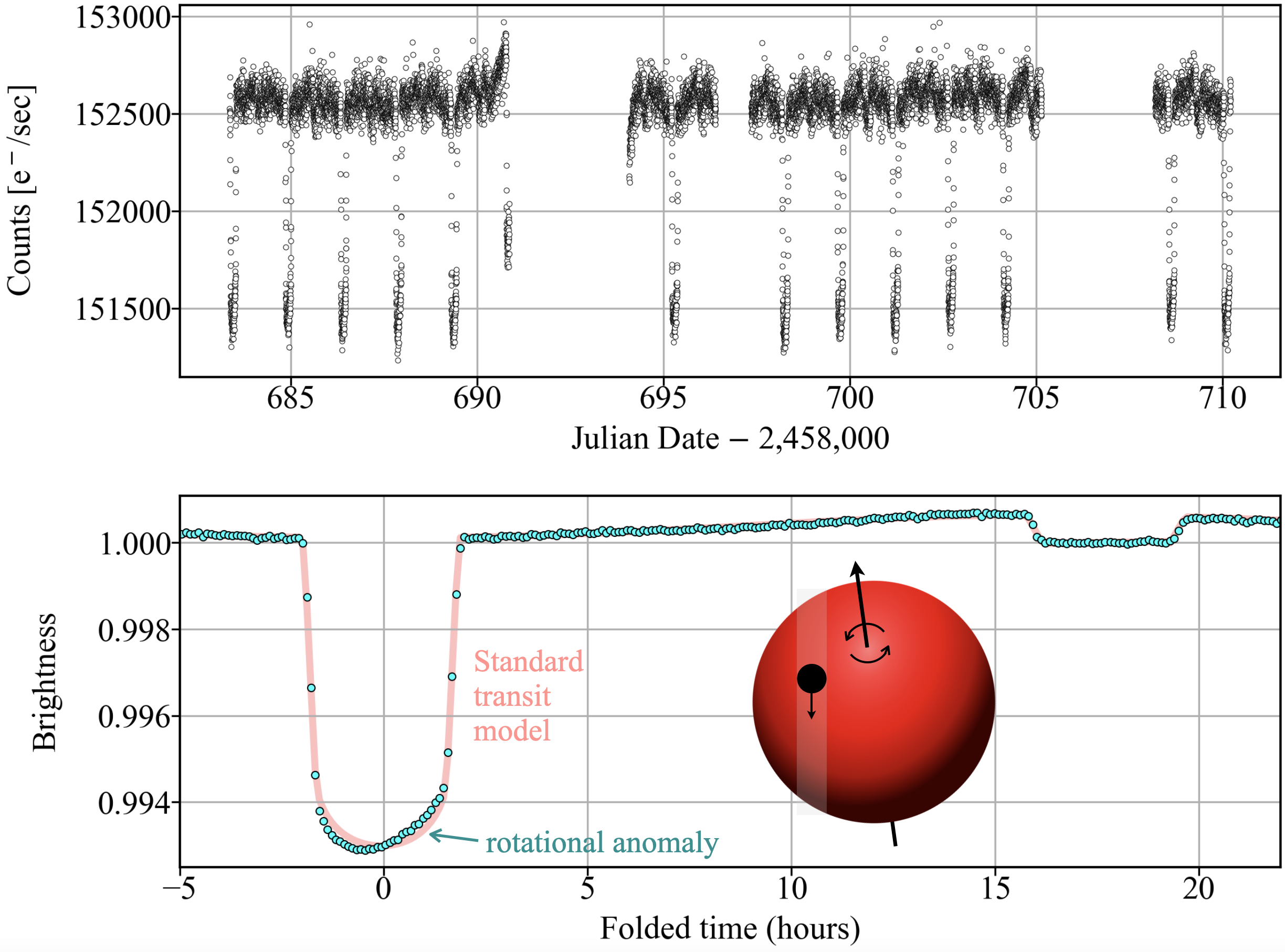}}
\caption{
{\bf An exemplary TESS light curve: the hot Jupiter KELT-9\,b}.
{\it Top.}---Data spanning one sector.
The two large gaps occurred
when the data quality was compromised by
stray light from the Earth. The smaller gap in the middle was the regularly scheduled interruption
of observations to allow TESS to
transmit data to the Earth.
{\it Bottom.}---Phase-averaged light curve based on five
sectors of data.
The enhanced signal-to-noise ratio
makes it easy to
detect the occultations that occur when
the planet goes behind the star.
The precision is also high enough
to detect an asymmetry in the transit light curve.
The symmetry is broken by two effects:
a misalignment between the
planet's orbit and the star's equatorial plane;
and the star's rapid rotation, which
causes its equatorial zone
to be lifted to higher elevation and become
cooler and fainter
\citep{Ahlers+2020}.
\label{fig:example_lightcurve}
}
\end{figure}

\runinhead{Transiting planet detection} The TESS Science Team
searches light curves for sequences of
periodic dips that may indicate the presence of transiting
planets. When a sequence passes the prescribed tests for statistical
significance and reliability,
the source is designated a
TESS Object of Interest (TOI) and assigned a serial number.
As a measure of photometric performance,
Figure~\ref{fig:tois} shows the
amplitudes of the detected transit signals
as a function of apparent magnitude
of the host star.
Figure~\ref{fig:example_lightcurve} provides an example of a TESS light curve.

\runinhead{Follow-up observations}  The TESS Follow-up Observing Program (TFOP) comprises hundreds of astronomers who coordinate and conduct observations to validate and confirm transiting planet candidates. This includes imaging observations with higher angular resolution and spectroscopic observations for stellar characterization and planetary mass measurement. TFOP is open to any interested astronomer willing to adhere to the policies regarding data sharing and publication. ExoFOP-TESS is a publicly accessible website created and maintained by the NASA Exoplanet Science Institute, providing astronomers with access to a voluminous archive of TESS follow-up data, including TFOP data.

\runinhead{Accessing TESS data} Images, light curves, and other
data products are publicly available on
the Mikulski Archive for Space Telescopes (MAST)
without any proprietary period.
An open-source Python package
called {\tt lightkurve}
makes it easy to download and manipulate TESS data.
NASA also sponsors a Guest Investigator program, allowing
US-based astronomers to
request funding for TESS-based research, and
allowing any astronomer to 
request observations of targets with 120- or 20-second cadence.
Short-cadence observations can also be requested
by appealing for Director's Discretionary Time.


\section{Contributions to exoplanet science}
\label{sec:surveys}

\begin{figure}[ht!]
\centerline{\includegraphics[width=11.75cm]{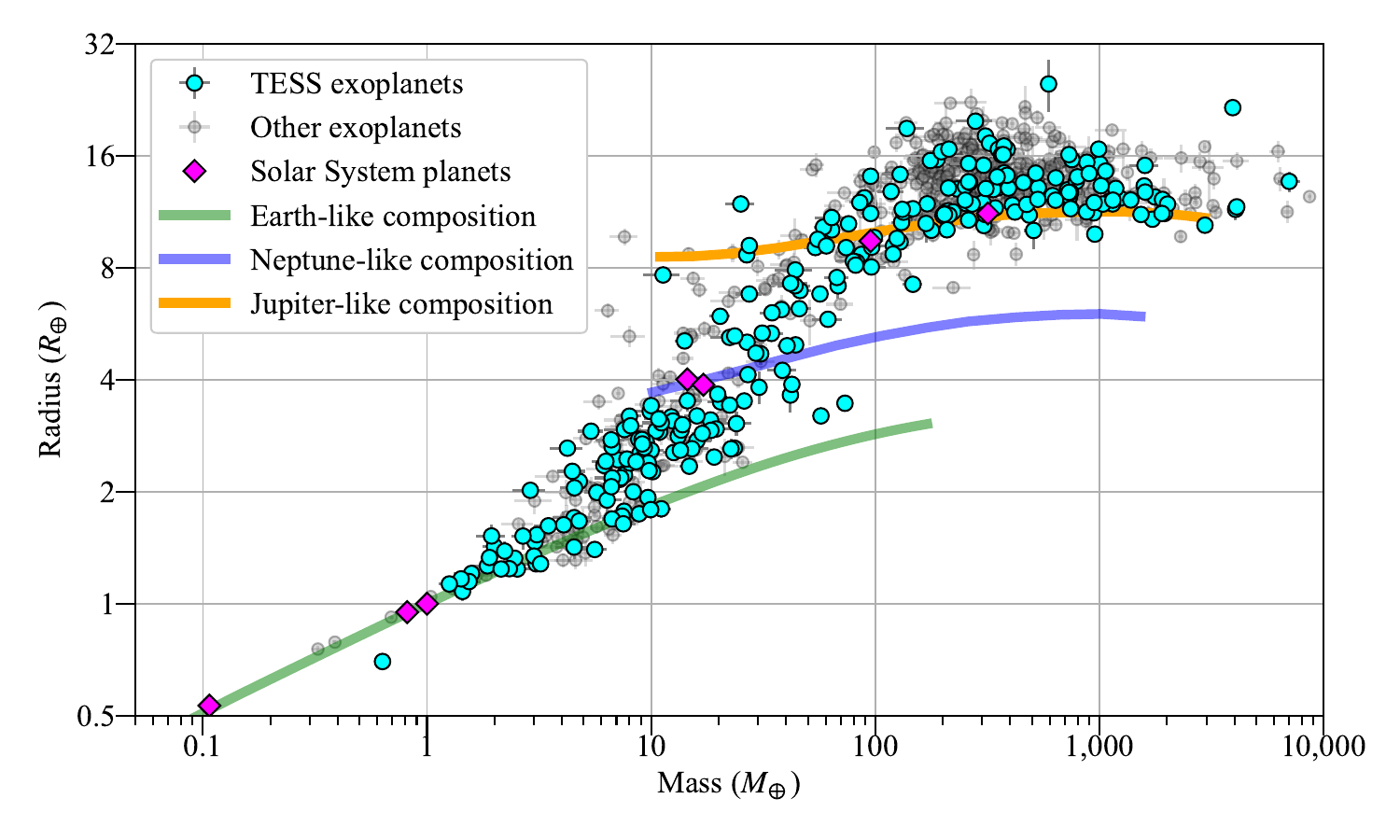}}
\vskip -0.15in
\caption{
{\bf Masses and radii} of transiting planets
for which both quantities have been measured with
a precision of 20\% or better. Light blue points
are for TESS planets, gray points are planets from
other surveys, and magenta diamonds
are for Solar System planets. The data are
from the NASA Exoplanet Archive. The curves
are approximate theoretical mass-radius relationships
from \cite{Fortney+2007} and \cite{Zeng+2019}.
\label{fig:mr}
}
\end{figure}

\runinhead{Masses and radii} Since the startling discovery
that at least a third of Sun-like stars host
systems of several planets orbiting within 1~AU and
ranging in size between Earth and Neptune
\citep{Lovis+2009,Latham+2011,Howard+2012,Fulton+2017},
the exploration
of ``super-Earths'' and ``mini-Neptunes'' has been
a high priority.
TESS has identified several hundred
such planets that are well-suited for mass measurement using the Doppler method. Accurately determining a planet's mass is crucial for inferring its composition and internal structure, as well as for interpreting data from atmospheric spectroscopy.
Figure~\ref{fig:mr} summarizes TESS's contribution to this endeavor. Even when considering only TESS planets,
one can see the progression in size and mass
from rocky Earth-sized planets
to inflated hot Jupiters.

\runinhead{JWST targets} One of TESS's intended roles 
was to act as a finderscope for JWST, by locating
the brightest and most
rewarding targets for atmospheric
spectroscopy.
As of 2024, TESS is responsible for having discovered about three-quarters of the transiting planets smaller than Neptune
for which JWST observations have been conducted or scheduled.
In its first three years of operations, JWST observed dozens of TESS planets, including:
\begin{itemize}

\item {\bf HIP\,67522\,b}, a Jupiter-sized planet with an orbital period of 7 days
and an age of only 17\,Myr
\citep{Rizzuto+2020}. Based on the strength of water
and carbon dioxide features in the JWST spectrum,
\cite{Thao+2024} found the planet's atmospheric
scale height to be several
thousand kilometers, much larger than
Jupiter's atmospheric scale height ($\sim$30\,km) and implying a
mean density lower than 0.1~g/cm$^3$.
Evidently, HIP\,67522\,b is more accurately classified as a ``super-puff'' planet than a hot Jupiter.

\item {\bf TOI-270\,d}, a $2.2\,R_\oplus$ planet orbiting a 0.38\,$M_\odot$ star \citep{Gunther+2019}.
The JWST spectrum obtained by
\cite{Benneke+2024} revealed absorption features from 
the volatile gases methane, carbon dioxide, and water. 
The inferred mean molecular weight of the upper atmosphere was $5.5\pm 1.2$~amu, lower than that of the volatile gases and higher than that of hydrogen and helium gas,
leading the investigators to propose that all of those
gases exist in a homogeneous mixture.

\item {\bf GJ\,367\,b}, a planet with a radius of
$0.70\,R_\oplus$ and an exceptionally high density of
$10.2\pm 1.3$~g/cm$^3$ orbiting
a 0.45\,$M_\odot$ star with a period of only 7.7~hours \citep{Lam+2021, Goffo+2023}. JWST observations covering a complete orbit failed to detect any hint of an atmosphere, supporting the notion that GJ\,367\,b and similar planets
are airless ``lava worlds'' \citep{Zhang+2024_GJ367}.
Earlier, a similar conclusion
was reached by \cite{Kreidberg+2019}
based on Spitzer observations of the TESS planet
LHS\,3844\,b \citep{Vanderspek+2019}.

\end{itemize}

\runinhead{Young planets} Many important processes
occur within the first $10^8$
years of a planetary system's life.
For example, high-energy radiation from a young star can strip hydrogen and helium
gas from the outer atmospheres of small planets.
As another example, the theories of hot Jupiter
formation make distinct predictions regarding their prevalence around young stars:
the high-eccentricity migration theory
predicts a low abundance,
while disk-driven migration theory
posits they should be equally common, if not moreso,
than around mature stars.
However, detecting young planets is difficult because young stars are both rare and photometrically variable. TESS helps by searching the entire sky and providing precise, well-sampled light curves, which are essential for assessing and mitigating stellar variability \citep[see, e.g.,][]{Mann+2020,Vach+2024}.
In addition to the low-density planet HIP\,67522\,b mentioned earlier, some notable
TESS discoveries are
\begin{itemize}

\item {\bf AU Microscopii}, the brightest pre-main-sequence star known to have transiting planets \citep{Plavchan+2020, Gilbert+2022}. The system is particularly interesting because it has a resolved edge-on debris disk, and the planetary orbits are aligned with the disk.

\item {\bf DS Tucanae}, a 45\,Myr old solar-mass star with a 5.7-$R_\oplus$
transiting planet in an 8-day orbit \citep{Newton+2019}.
The system is close enough to observe X-ray flares \citep{Pillitteri+2022}
and bright enough for the star’s obliquity to be measured using the Rossiter-McLaughlin effect \citep{Zhou+2020}.

\item {\bf HD\,114082}, a 15\,Myr old star
located in the lower Centaurus Crux association
with an edge-on debris disk. TESS observed
a transit of a ``super-Jupiter'' of mass
8~$M_{\rm Jup}$ on a 110-day orbit \citep{Zakhozhay+2022}.
This is one of very few
young planets for which the mass, radius, and age are all well-determined and can be compared to evolutionary models.

\item {\bf TOI-1136}, a Sun-like star 700 million years old
with six transiting planets forming a ``resonant chain'' \citep{Dai+2023}.
Starting from the innermost planet and proceeding
outward, the orbital period ratios are nearly
3:2, 2:1, 3:2, 7:5, and 3:2.
Although strongly resonant planetary systems
were already known to exist
from the Kepler mission, TESS data were used to show that they are
more common around younger stars, suggesting
that many planetary systems form in resonant chains
that gradually break over hundreds of millions of years \citep{Dai+2024}.

\end{itemize}

\noindent In addition to detecting transiting planets around known
young stars, TESS can confirm the youth of a population of stars by measuring their rotation periods and amplitudes of photometric variability. 
For example, \cite{Curtis+2019} used
TESS data to confirm the existence of the Psc-Eri stream of stars, revising its estimated age downward
to 120~Myr. This younger age establishes the Psc-Eri stream as a benchmark for stellar astrophysics, comparable to the Pleiades.
Another example is the open cluster $\delta$\,Lyr, which lies within the original Kepler mission's field of view. TESS measurements of rotation periods confirmed the cluster's existence and determined its age to be 38 Myr \citep{Bouma+2022}.

\runinhead{Mine is brighter} A popular genre of TESS
papers is ``We used TESS data to find the {\tt [superlative]} 
known {\tt [planet type]}''.
Keeping in mind that all such records are meant
to be broken, some representative examples are:
\begin{itemize}

\item {\bf HD\,110067}, the brightest known
star with six transiting planets in a chain of mean-motion resonances \citep{Luque+2023}.

\item {\bf LTT\,1445\,A}, the nearest
M dwarf star known to have a transiting planet \citep{Winters+2019}.

\item {\bf Gliese 12\,b}, the nearest known transiting temperate Earth-sized planet \citep{Dholakia+2024,Kuzuhara+2024}.

\item {\bf HD\,63433\,d}, the smallest confirmed transiting
planet younger than 500 Myr, and the nearest
young Earth-sized planet \citep{Capistrant+2024}.

\item {\bf TIC 172900988} and {\bf TOI-1338}, the two brightest
eclipsing binaries with transiting circumbinary planets \citep{Kostov+2020,Kostov+2021}.

\end{itemize}

\runinhead{Statistical surveys} Although TESS cannot yet match Kepler's sensitivity to signals of small amplitudes ($<$10$^{-4}$) and long periods ($>$27~days),
TESS compensates by observing the entire sky and collecting data for a much larger number of stars. Consequently, for rare objects with relatively large-amplitude signals, TESS offers better data for population statistics. Some examples are:
\begin{itemize}

\item {\bf Hot Jupiters.} TESS data allow for the unification
and expansion of 
the earlier ground-based wide-field surveys for hot Jupiters,
which has led to a magnitude-limited sample
of approximately 400 hot Jupiters around
Sun-like stars \citep[see, e.g.,][]{Zhou+2019,Yee+2023, Schulte+2024}.
Surveys for hot Jupiters have also been undertaken
for massive stars \citep{BeleznayKunimoto2022}
and low-mass stars \citep{Gan+2023_HJOccurrence,Kanodia+2024}.

\item {\bf Nearby companions of hot Jupiters.}
TESS has found at least seven systems where a hot Jupiter
has a smaller planetary companion on an interior
orbit \cite[see, e.g.,][]{Hord+2022,Korth+2024}. These findings
challenge the prevailing notion that hot Jupiters generally lack close companions, a trend previously interpreted as evidence of high-eccentricity migration.

\item {\bf Brown dwarfs.} 
Since the discovery 
of the first brown dwarf in 1995, only about 50 transiting or eclipsing
brown dwarfs have been found. About two-thirds of them have been
found using TESS data \cite[see, e.g.,][]{Carmichael+2023,Henderson+2024}.

\item {\bf Terrestrial planets orbiting mid-to-late M dwarfs.}
\cite{MentCharbonneau2023} used TESS data to measure the
occurrence of planets with sizes 0.5--2\,$R_\oplus$ and periods
shorter than a week around stars of masses
0.1--0.3\,$M_\odot$.
complementing earlier work on early M dwarfs
based on Kepler data.


\end{itemize}

\runinhead{Transit timing.} Each new TESS observation of a transit
refines our knowledge of the planet's period and improves our
ability to predict future transit times.
This improved accuracy, in turn, increases the efficiency of
follow-up observations with costly facilities such
as JWST. The precise photometry and extensive time baseline of the
TESS survey also provides leverage for the detection of
any changes in the orbital period \citep[][]{IvshinaWinn2022},
such as those expected from
tidal orbital decay \citep{Wong+2022, Vissapragada+2022}
or the influence of wide-orbiting companion planets
\citep{Bouma+2020_WASP4}.

\runinhead{`Impossible' planets.} TESS data have been used
to identify a few stars with planets
where one might never have expected to find them:
\begin{itemize}

\item {\bf 8\,Ursae Minoris} has a Doppler-discovered
giant planet in a circular orbit of radius 0.2\,AU.
Observations of the star's optical variability by TESS revealed it to be a helium-burning star, implying that a few hundred million years ago, the star was a red giant star with a radius
extending beyond the planet's current orbit \citep{Hon+2023}.
How did the planet survive being engulfed
by the star? A possible solution to this riddle
is that the planet used to be a circumbinary planet.
In this scenario, the primary star's expansion
was interrupted by a merger with
the secondary star, creating
a helium-burning merger product.

\item {\bf WD\,1856+534} is a white dwarf with a transiting
Jupiter-sized companion in a 0.02\,AU orbit \citep{Vanderburg+2020}.
How did this planet attain such a tight
orbit around an erstwhile red giant star? Perhaps the planet
underwent high-eccentricity migration,
in a post-main-sequence variation of the same
theory that has been proposed to explain the presence of hot Jupiters around
main-sequence stars.

\end{itemize}

\begin{figure}[p]
\includegraphics[width=11.75cm]{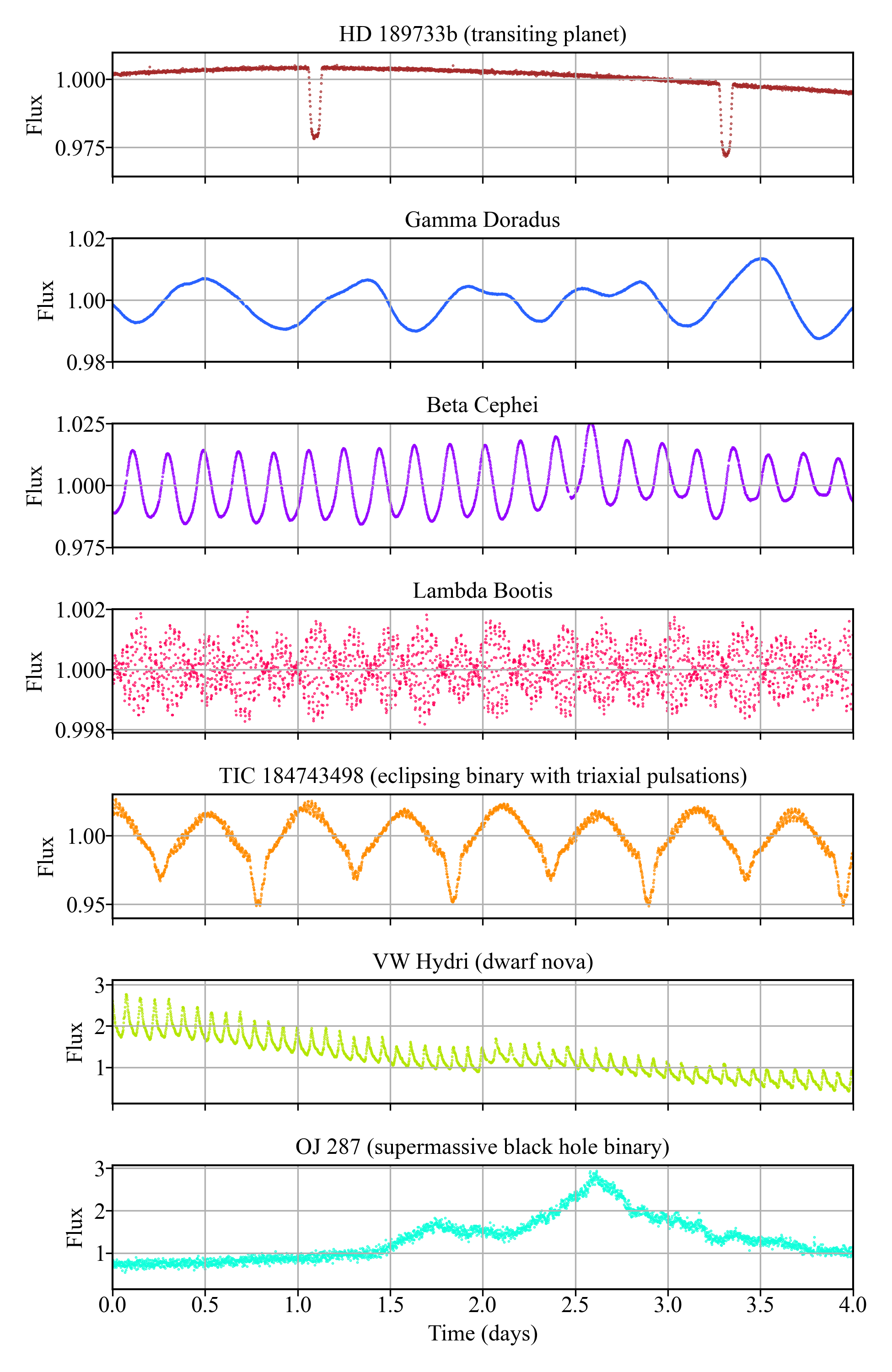}
\caption{
{\bf Four days of TESS photometry} for seven
types of astronomical sources: a hot Jupiter;
the eponymous stars 
Gamma Doradus,
Beta Cephei, and Lambda Bootis that
define categories of pulsating variables;
a `triaxial pulsator' in which a star's usual
pulsations are reshaped by tides from a close companion \citep{Handler+2020,Zhang+2024_Triaxial};
a dwarf nova exhibiting frequent eruptions;
and an active galactic nucleus suspected
of being a binary supermassive black hole \citep{Kishore+2023}.
\label{fig:gallery}
}
\end{figure}


\section{Contributions to other areas of astronomy}
\label{sec:stellar}

Although this article is part of the Handbook on {\it Exoplanets},
no dicussion of TESS would be complete without at least a brief
overview of its contributions to other areas
of astronomy.

\runinhead{Stellar astrophysics.} The majority of TESS publications are not about exoplanets --- they are actually about stars.
TESS is re-energizing the revolution in stellar astrophysics that began with the CoRoT and Kepler missions, particularly in
the realm of asteroseismology \citep{Kurtz2022}. 
TESS data have been used to study the variability
of stars across the entire Hertzsprung-Russell diagram, encompassing timescales ranging from a few minutes to months, and amplitudes ranging from micro-magnitudes to factor-of-fifty flares. 
Some highlights are:

\begin{itemize}

\item Detection of oscillations in 150{,}000 red giants, an order of magnitude more than detected by the Kepler mission, providing a nearly all-sky map of stellar masses throughout the galaxy \cite{Hon+2021}.

\item A catalog of a million stellar flares, enabling the first population-level analysis of flaring as a function of stellar mass
\cite{Feinstein+2022}.

\item Discovery of young $\delta$ Scuti pulsators with regularly spaced high frequencies. This allowed the first systematic identification of individual pulsation modes in these stars \citep{Bedding+2020}.

\item Identification of the brightest known detached eclipsing binary \citep{Bedding+2019},
publication of a catalog of 4{,}000 eclipsing binaries \citep{Prsa+2022},
and publication of a separate catalog
of 15{,}000 candidate
`ellipsoidal' binaries
(identified from photometric modulations
attributed to the tidally
stretched shapes of the stars).

\item A series of papers published in {\it The Observatory} by J.~Southworth in which the masses and radii of the brightest eclipsing binaries
are lovingly measured
by combining exquisite TESS light curves with archival spectroscopy. In many
cases, the absolute dimensions of the
stars are determined to better than a few percent \citep[see, e.g.,][]{Southworth+2021,Southworth+2024}.

\item Discovery of ubiquitous stochastic variability in O and B-type stars, which is likely a signature of internal gravity waves, which bear information about mixing processes in the massive stars that are progenitors of neutron stars and black holes \citep{Bowman+2019,Shen+2023}.

\item Determination of the age of the Gaia-Enceladus merger with the Milky Way galaxy, based on asteroseismology of the nearby naked-eye star Nu Indi \citep{Chaplin+2020}.

\end{itemize}

\runinhead{Asteroids} Most main-belt asteroids vary by 0.1 mag or more on the hours-to-days timescales accessible to TESS. The period, amplitude, and shape of the light curve can be used to measure the spin rate and direction, and sort asteroids into families with different spin properties, thereby elucidating their origin. TESS data were used
to construct $\sim$$10^4$ asteroid light curves, primarily main-belt asteroids and Jovian Trojans \citep{Pal+2020}.

\runinhead{Comets} TESS observations of Comet 46P/Wirtanen probably represent the best temporal coverage of a cometary outburst ever recorded \citep{Farnham+2021}. Further afield, transits of presumed `exocomets' were detected
across the face of Beta Pictoris \citep{Zieba+2019,Lecavelier_des_Etangs+2022}, confirming earlier
evidence for comets in that system based on spectroscopic
transients \citep{Ferlet+1987}.

\runinhead{Galactic transients} TESS data have been used to
search for photometric periodicities and study period variations
in the light curves of nova outbursts, cataclysmic variables, and X-ray binaries (see, e.g., \citealt{Rawat+2021, Ilkiewicz+2021, Bruch2024} and the second lowest panel of Figure~\ref{fig:gallery}).

\runinhead{Active galactic nuclei} Optical variability studies with TESS data have been used to identify active galactic nuclei
\citep{Treiber+2023}, detect periodicities \citep{Kishore+2023},
and observe
flares (\citealt{Kishore+2024}; see the lowest panel of Figure~\ref{fig:gallery}).

\runinhead{Extragalactic transients} TESS 
provides useful light curves for supernovae with peak magnitudes brighter than about 20. \cite{Fausnaugh+2023} used about
300 Type Ia supernova light curves from TESS
to search for evidence of companion star interactions that
are predicted in single-degenerate explosion models.
TESS data have also been used to
study unusually bright tidal disruption events \citep{Holoien+2019},
possibly repeating tidal disruption events \citep{Payne+2021},
and optical emission from gamma-ray bursts
\citep{Fausnaugh+2023_grb, Roxburgh+2024}.

\runinhead{Cosmology} The author is not aware of any contributions TESS has made to the measurement of cosmological parameters, the detection of dark matter, or the unraveling of the mystery of dark energy. Well, nobody's perfect.


\section{Cross-References}
\begin{itemize}
\item{PLATO}
\item{Planet Occurrence from Doppler and Transit Surveys}
\item{Space Missions for Exoplanet Science: Kepler/K2}
\item{Space Missions for Exoplanet Science: CoRoT}
\end{itemize}

\section{Acknowledgements} The success
of the TESS mission is due to the effort of
thousands of people --- managers, engineers, scientists, and administrators, among others --- who are too numerous to be named here but are deeply appreciated.
The author is grateful to George Ricker and David
Latham for the invitation to join the TESS team in 2006.
Simon Albrecht, Luke Bouma, Fei Dai,
Daniel Huber, David Latham, Roland Vanderspek, and Eritas Yang provided helpful suggestions on the manuscript.


\bibliographystyle{spbasicHBexo}  
\bibliography{references} 

\end{document}